%% file: cavitation.tex
\documentclass[
    aip,
    pof,
    reprint]{revtex4-1}

\input{preamble}

\myexternaldocument[si-]{cavitation_si}

\makeatletter
\def\@email#1#2{%
 \endgroup
 \patchcmd{\titleblock@produce}
  {\frontmatter@RRAPformat}
  {\frontmatter@RRAPformat{\produce@RRAP{*#1\href{mailto:#2}{#2}}}\frontmatter@RRAPformat}
  {}{}
}%
\makeatother

\begin{document}

\preprint{AIP/123-QED}
\title{\manuscripttitle}
\input{authors}

\input{main/sections/abstract}

\maketitle

\input{main/sections/introduction}
\input{main/sections/methods}
\input{main/sections/results}
\input{main/sections/conclusion}
\input{main/sections/supplementary}
\input{main/sections/acknowledgments}
\input{main/sections/data_availability}

\bibliography{bibliography/matej, bibliography/philip, bibliography/roland}

\end{document}

\title{\manuscripttitle:~Supplementary~Material}

\input{authors}

\maketitle
\tableofcontents

\input{si/sections/polarity_angle}
\input{si/sections/fitting_prefactor}
\input{si/sections/kappa_0_2D_exponent}
\input{si/sections/cavitation_pressure_tau}
\input{si/sections/theta_star}

\bibliography{bibliography/matej, bibliography/philip, bibliography/roland}

%% file: preamble.tex
\usepackage[english]{babel}
\usepackage{color}
\usepackage{comment}
\usepackage[utf8]{inputenc}
\usepackage[T1]{fontenc}
\usepackage[fleqn]{mathtools}
\usepackage{sidecap}
\usepackage{xr}

\usepackage{hyperref}

\newcommand{\manuscripttitle}
    {Water cavitation  results from the kinetic competition of bulk, surface and
    surface-defect nucleation events}

\addto\extrasenglish{}
\addto\extrasenglish{}

\makeatletter
\newcommand*{\addFileDependency}[1]{% argument=file name and extension
\typeout{(#1)}
\@addtofilelist{#1}
\IfFileExists{#1}{}{\typeout{No file #1.}}
}\makeatother

\newcommand*{\myexternaldocument}[2][]{%
\externaldocument[#1]{#2}%
\addFileDependency{#2.tex}%
\addFileDependency{#2.aux}%
}

\newcommand{\rev}[1]{\textcolor{black} {#1}}

%% file: authors.tex
\author{Philip Loche}
\affiliation{Laboratory of Computational Science and Modeling, IMX,
    École Polytechnique Fédérale de Lausanne, 1015 Lausanne, Switzerland}
\affiliation{Fachbereich Physik, Freie Universität Berlin,
    14195 Berlin, Germany}
\author{Matej Kanduč}
\affiliation{Department of Theoretical Physics, Jožef Stefan Institute,
    1000 Ljubljana, Slovenia}
\author{Emanuel Schneck}
\affiliation{Physics Department, Technische Universität Darmstadt,
    64289 Darmstadt, Germany}
\author{Roland R.\ Netz}
\email{rnetz@physik.fu-berlin.de}

%% file: main/sections/abstract.tex
\begin{abstract}
  Water at negative pressures can remain in a metastable state for a surprisingly long
  time before it reaches equilibrium by cavitation, i.e. by the formation of vapor
  bubbles. The wide spread of experimentally measured cavitation pressures depending on
  water purity, surface contact angle and surface quality implicates the relevance of
  water cavitation in bulk, at surfaces and at surface defects for different systems. We
  formulate a kinetic model that includes all three different cavitation pathways and
  determine the needed nucleation attempt frequencies in bulk, at surfaces and at
  defects from atomistic molecular dynamics simulations.
  Our model reveals that cavitation occurs in pure bulk water only for defect-free
  hydrophilic surfaces with wetting contact angles below 50° to 60° and at pressures of
  the order of \textminus100\,MPa, depending only slightly on system size and
  observation time. Cavitation on defect-free surfaces occurs only for higher contact
  angles, with the typical cavitation pressure rising to about \textminus30\,MPa for
  very hydrophobic surfaces. Nanoscopic hydrophobic surface defects act as very
  efficient cavitation nuclei and can dominate the cavitation kinetics in a macroscopic
  system. In fact, a nanoscopic defect that hosts a pre-existing vapor bubble can raise
  the critical cavitation pressure much further. Our results explain the wide variation
  of experimentally observed cavitation pressures in synthetic and biological systems
  and highlight the importance of surface and defect mechanisms for the nucleation of
  metastable systems.
\end{abstract}

%% file: main/sections/introduction.tex
\section{Introduction}
When water is brought into a metastable state by either heating or pressure reduction,
it will eventually reach equilibrium by cavitation, i.e., by the nucleation of vapor
bubbles. Cavitation and the collapse of cavitation bubbles are central to applications
such as inkjet printing, ultrasonic cleaning, noninvasive destruction of kidney stones
by acoustic shockwaves, sonochemistry,
sonoluminescence~\cite{coleman1989survey,caupin2006cavitation} \rev{and for the
investigation of geothermal processes\cite{fall_effect_2009}.} In contrast, the collapse
of hydrodynamically produced cavitation bubbles has adverse effects on hydraulic
systems, valves, and propeller blades, causing wear and
erosion~\cite{dular2004relationship, reuter2022cavitation, adhikari2015mechanism},
\rev{while the creation of cavitation bubbles may impact the performance of
hydrofoils~\cite{custodio_cavitation_2018}.}
In biology, where temperature is fixed by the environment, metastable water is primarily
produced by pressure reduction. The collapse of these cavitation bubbles provides
catapult-like mechanisms in ferns~\cite{noblin2012fern} and allows snapping shrimp to
stun their prey \cite{LohseScience2000}. Conversely, cavitation limits the negative
pressure produced by octopus suckers~\cite{smith1991negative} and must be avoided in the
ascending sap of tall trees, where significantly negative pressures are encountered, in
Importantly, metastable water at negative pressures can persist over astonishingly long
times and cavitation occurs at experimentally relevant time scales only when the
pressure sinks below a threshold cavitation pressure $p_\mathrm{cav}$.
%, which can be theoretically as low as \textminus100\,MPa~\cite{caupin2006cavitation}.
The large variation of $p_\mathrm{cav}$ among different systems and its dependence on
surface properties suggests that cavitation in bulk water (i.e., homogeneous nucleation)
competes with cavitation on surfaces and with cavitation at surface defects (i.e.,
heterogeneous nucleation). Understanding the kinetic competition between these three
different cavitation pathways is important for all the above-mentioned applications and
is the central theme of this paper.

% CNT and experiments
Cavitation is a typical nucleation phenomenon; according to Classical Nucleation Theory
(CNT), its kinetics is determined by the thermally activated stochastic crossing of a
free energy barrier characteristic of the critical nucleus
size~\cite{debenedetti1996metastable}. CNT predicts that clean water at room temperature
can sustain negative pressures considerably below \textminus100\,MPa before bulk
cavitation sets in~\cite{fisher_fracture_1948, caupin2005liquid, caupin2006cavitation,
caupin2013stability, azouzi2013coherent}. This very low cavitation pressure predicted
for clean bulk water is largely due to the high water surface tension. Experimentally,
cavitation pressures below \textminus100\,MPa have only been reached with ultra-clean
water in microscopic quartz cavities~\cite{zheng_liquids_1991, alvarenga1993elastic,
azouzi_coherent_2013}. In typical experiments with hydrophilic surfaces and using
rigorous purification and degassing methods, cavitation pressures of the order of
$p_\mathrm{cav}\approx -30$\,MPa~\cite{caupin2006cavitation, caupin2013stability,
caupin2015escaping} are reached, significantly above the CNT results for clean bulk
water. However, with untreated ordinary water in standard containers, the cavitation
pressure is positive and typically only slightly below the saturated water vapor
pressure of about 2\,kPa.
The consensus in the field is that surface cavitation at container walls, microscopic
impurities~\cite{jones_bubble_1999, marschall_cavitation_2003, sear2007nucleation,
morch2007reflections, tsuda_study_2008, gross2017diffusion, gao2021effects}, surface
imperfections, or crevices hosting pre-existing vapor or gas
bubbles~\cite{harvey_bubble_1944, atchley_crevice_1989, borkent2009nucleation,
pfeiffer2022heterogeneous} act as cavitation nuclei. These nuclei promote cavitation and
contribute to deviations in cavitation pressures from those predicted by CNT for clean
bulk water.

\rev{These experimental complications make theoretical and simulation approaches
  essential. Molecular dynamics (MD) simulations with all-atom force fields have proven
  to be powerful tools for investigating cavitation
  processes~\cite{menzl_molecular_2016, p2023water,min2019bubbles, xie2022study}. Recent
  MD studies demonstrated that while dissolved gas has a negligible effect on cavitation
  thresholds~\cite{zhou2019molecular, kanduc_cavitation_2020}, gas nanobubbles can
  significantly promote cavitation~\cite{dockar_mechanical_2019, gao2021effects}.
  Indeed, our recent MD study demonstrated that nonsoluble impurities in the form of
  nanodroplets suspended in water considerably increase cavitation
  pressures~\cite{Sako2024droplet}. Another simulation study has shown that solid
  particles in water can disrupt the hydrogen-bond network and facilitate cavitation in
  small simulation volumes~\cite{li2019cavitation}. Despite these advances, the
  influence of specific surface properties on cavitation in macroscopic systems remains
  poorly understood, and a unified framework encompassing bulk, surface, and
defect-induced cavitation is still missing.}

\rev{In this work, we bridge this gap by formulating a kinetic cavitation model that
  combines MD simulations with the CNT framework. With attempt rates obtained from
  simulations, our kinetic model predicts the cavitation pressure, $p_\mathrm{cav}$, of
  water confined in a cubic container, with walls characterized by the water contact
  angle. The model also incorporates various surface defect patches with distinct
  wetting properties. We find that for highly hydrophilic, defect-free surfaces,
  cavitation occurs in bulk water at pressures consistent with CNT predictions. As the
  surface contact angle increases beyond a critical threshold of 50° -- 60°, cavitation
  shifts to the surface. Finally, for surfaces with hydrophobic defects our
  comprehensive model provides a unified framework to understand and predict cavitation
  behavior in diverse practically relevant scenarios.}

%% file: main/sections/methods.tex
\section{Materials and Methods}
\subsection*{Simulation model}
\begin{figure}
  \includegraphics[width=0.8\columnwidth]{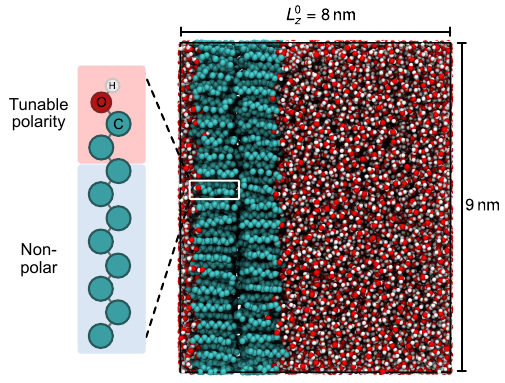}
  \caption{\rev{Simulation setup consisting of 800 surface molecules (hydroxylated
      alkanes) assembled into two self-assembled monolayers and solvated by 16,353 water
      molecules. Each surface molecule consists of an alkyl chain terminated by a
      modified hydroxyl group with its partial charges scaled by a factor $\alpha$.
      Simulation box (black frame) is replicated in all three directions via periodic
      boundary conditions. The box length for the initial pressure of
      $-30\,\mathrm{MPa}$ in the normal direction is $L_z^0=8\,\mathrm{nm}$.}}
  \label{fig:system}
\end{figure}
The simulation setup follows our previous approach \cite{kanduc2014attraction,
kanduc_atomistic_2017} \rev{and is shown in \autoref{fig:system}}. We model a planar
surface composed of two self-assembled monolayers, each consisting of ten-carbon-atom
alkyl chains terminated by polar a hydroxyl ($\mathrm{OH}$) head group. These
hydroxylated alkanes are arranged on a hexagonal lattice with an areal density of
$4.3\,\mathrm{nm^{-2}}$. To stabilize the structure, we apply harmonic restraints to the
second carbon atom from the head group with strengths of $k_x = k_y =
500\,\mathrm{(kJ/mol)\,nm^{-2}}$ in the lateral directions, and to the tenth (terminal)
carbon atom with $k_x = k_y = 10\,\mathrm{(kJ/mol)\, nm^{-2}}$ laterally and $k_z =
100\,\mathrm{(kJ/mol)\, nm^{-2}}$ in the normal direction. The surface polarity and
thereby the surface contact angle is controlled by scaling the partial charges of the
head groups with a factor $\alpha$, ranging from 0 to 1. For $\alpha = 0$, the surface
is non-polar with head groups resembling methylated termini, resulting in a contact
angle of $\theta=135^\circ$. For $\alpha = 1$, the surface is fully polar, mimicking
hydroxyl head groups. \rev{In the Supplementary Material, section S1, we present a
phenomenological fit function that maps the surface polarity $\alpha$ to the contact
angle $\cos\theta$.} The periodic simulation box, with initial dimensions
$9\,\mathrm{nm}\times10.4\,\mathrm{nm}\times8\,\mathrm{nm}$ at a pressure of
$-30\,\mathrm{MPa}$, contains 16,353 water molecules and 800 hydroxylated alkanes.

We use united-atom parameters from the GROMOS force field for the
alkanes~\cite{oostenbrink_biomolecular_2004, kanduc_atomistic_2017} and the rigid SPC/E
water model~\cite{berendsen_missing_1987}. Classical non-polarizable molecular dynamics
simulations are performed with the GROMACS package~\citep{abraham_gromacs:_2015} using a
time step of $2\,\mathrm{fs}$.
%The total simulation length in the pressure ramp the protocol depends on the pressure
%rate $\dot p$.
A plain cutoff of $0.9\,\mathrm{nm}$ is used for short-range Lennard-Jones interactions.
Electrostatics is treated with the Smooth Particle--Mesh Ewald method, using a
$0.9\,\mathrm{nm}$ real-space cutoff. A temperature of 300~K is controlled using the
velocity rescale thermostat with a stochastic factor~\cite{bussi_canonical_2007} and a
time constant of $0.5~\mathrm{ps}$. Pressure is controlled by the Berendsen barostat
with a time constant of $1\,\mathrm{ps}$, suitable even for negative pressures because
of its efficiency and stability in box scaling under large pressure differences. We
previously verified~\cite{kanduc_cavitation_2020} that equivalent results are obtained
with the Parrinello--Rahman barostat, which more accurately reproduces volume
fluctuations. Reference coordinates for surface-atom restraints are scaled with the
coupling matrix of the pressure coupling.

\subsection*{Kinetic theory for pressure ramps}
To obtain the cavitation attempt frequencies, we use the pressure-ramp simulation
protocol~\cite{kanduc_cavitation_2020}, where the negative pressure decreases linearly
over time as $p(t)=\dot p t$ at a constant rate $\dot{p}<0$. To simulate pressure ramps,
we conduct a sequence of simulations at constant pressures, starting from an initial
pressure of $-30\,\mathrm{MPa}$. After each simulation, the final coordinates are used
to start the next simulation, with the pressure reduced by $0.1\,\mathrm{MPa}$. The
duration of each constant-pressure simulation depends on the pressure rate $\dot p$.  We
identify cavitation when the simulation box length $L_z$ increases by more than 50\,\%
of its initial value $L_z^0$. The corresponding ``dynamic'' cavitation pressure
$p^*_\mathrm{cav}$ follows from the solution of the linear cavitation rate equation as
\begin{align}\label{eq:p*cav}
  p^*_\mathrm{cav} = \dot{p} \int_0^\infty e^{-k_0 I(t)} \mathrm{d}t\,,
\end{align}
where
\begin{equation}
  I(t)=\int_0^t e^{-\beta G^*(t')}dt'\,.
\end{equation}
Here, $\beta=1/k_\mathrm{B}T$ is the inverse thermal energy, and $G^*(t)$ is the
time-dependent free energy barrier in the pressure-ramp protocol. According to CNT, this
barrier decays with time as $ G^*(t) \sim t^{-2}$.
%In this case, the integral $I(t)$ has a closed-form
%solution~\cite{kanduc_cavitation_2020} \begin{equation} I(t)= t
%\exp\left[-(\tau/t)^2\right]-\sqrt{\pi}\tau\,\textrm{erfc}(\tau/t) \end{equation} where
%$\textrm{erfc}(x)$ is the complementary error function.
The attempt frequencies for bulk and surface cavitation, $\kappa_\mathrm{3D}$ and
$\kappa_\mathrm{2D}$, are obtained from fits of \autoref{eq:p*cav} to the simulation
data; detail are given in the Supplementary Material, section S2.
%The kinetic prefactor for bulk water $\kappa_\mathrm{3D}=5\times
%10^{11}\,\mathrm{ns^{-1}nm^{-3}}$ is taken from our previous work
%\cite{kanduc_cavitation_2020}, based on 30--50 independent simulation runs at a
%constant pressure rate of $\dot{p}=-5\,\mathrm{MPa/ns}$.

%% file: main/sections/results.tex
\section{Results and Discussion}
%%%%%%%%%%%%%%%%%%%%%%%%%%%%%%%%%%%%%%%%%%%%%%%%%%%%%%%%%%%%%%%%%%%
\subsection{Kinetic model for cavitation in water confined by surfaces with defects}
% Vessel/Enclosure/Container \textbf{Competition between homogeneous and
% heterogeneouscavitation}
%
In the typical cavitation scenario, \rev{as studied in this work}, water is
confined by solid surfaces with a given contact angle that contain a finite number of
surface defects. Upon application of a negative pressure, cavitation can proceed in
bulk, at the defect-free surface parts or at the defects. Since we cannot presuppose the
prevalence of one cavitation pathway over the others, we write the total cavitation
rate as the sum of the bulk (3D), surface (2D) and defect (def) rates as
\begin{align}\label{eq:k_tot}
  k_\mathrm{tot} = k_\mathrm{3D} + k_\mathrm{2D}+ N_\mathrm{def} k_\mathrm{def} \, ,
\end{align}
where $N_\mathrm{def}$ is the number of identical surface defects. The model assumes that the
different cavitation pathways do not interfere with each other. %, which we confirm by our MD simulations.
It will turn out that depending on the surface and defect properties,
quantified by their respective contact angles, one cavitation pathway will always
dominate strongly over the other pathways, such that we can construct effective kinetic
phase diagrams that feature regions where bulk, surface or defect cavitation dominates.

%%%%%%%%%%%%%%%%%%%%%%%%%%%%%%%%%%%%%%%%%%%%%%%%%%%%%%%%%%%%%%%%%%%
\subsection{Cavitation in bulk}
CNT is a general framework for the description of
cavitation~\cite{debenedetti1996metastable}. For cavitation in bulk, one considers a
spherical vapor bubble in a liquid with a free energy given by~\cite{caupin2006cavitation,
azouzi2013coherent, menzl_effect_2016}
\begin{align}\label{eq:free_energy_3D}
  G_\mathrm{3D} = 4 \pi r^2 \gamma + \frac{4}{3} \pi r^3p\,,
\end{align}
where $\gamma$ is the liquid--vapor surface tension, $r$ the bubble radius and $p<0$ the
negative pressure in the liquid. The first term represents the bubble surface free
energy and the second term the work associated with the volume expansion. We have
neglected the saturated vapor pressure, which is irrelevant compared to the large
magnitude of cavitation pressures considered here. For simplicity, we have also
neglected curvature effects on the surface tension, which only introduce minor
corrections~\cite{menzl_molecular_2016, kanduc_cavitation_2020}. The interplay between
the two terms in \autoref{eq:free_energy_3D} creates a free energy barrier for $p<0$ of
\begin{align}\label{eq:G_3D}
  G_\mathrm{3D}^* = \frac{16 \pi }{3}\frac{\gamma^3}{p^2} \,,
\end{align}
reached at the critical bubble radius $r^*=-2\gamma/p$ defined by
$\mathrm{d}G_\mathrm{3D}/\mathrm{d}r |_{r=r*} =0$. If a bubble surpasses the critical
radius $r^*$, it will grow further and cause cavitation.

Cavitation can occur anywhere in bulk, its rate therefore is proportional to the liquid
volume $V$ and can be written as
\begin{align}\label{eq:k_3D}
  k_\mathrm{3D} = V \kappa_\mathrm{3D}\,
  e^{-\beta G^*_\mathrm{3D}}\,.
\end{align}
The cavitation time follows from the rate as $\tau=k_\mathrm{3D}
^{-1}$~\cite{hanggi1990reaction} and is the mean time one has to wait for cavitation to
occur. The exponential factor in \autoref{eq:k_3D} reflects the Arrhenius
law~\cite{hanggi1990reaction}.
The pre-exponential factor splits into the volume $V$ and the attempt frequency density
$\kappa_\mathrm{3D}$, which is an intensive property of the liquid that characterizes
the barrier-less cavitation rate. It thus constitutes the cavitation speed limit, i.e.,
the maximum rate at which cavitation occurs ub te absence of a barrier
\cite{speedlimitproteinfolding}. The attempt frequency density, $\kappa_\mathrm{3D}$,
accounts for viscosity and hydrodynamic memory effects in the surrounding liquid, as
well as for corrections to CNT, such as the curvature dependence of the interfacial
tension \cite{Bagchi2024}.
Given the exponential dependence of the cavitation time $\tau$ on $p$, it is preferred
to consider the inverse relation. For a cubic volume $V=L^3$, the cavitation pressure
follows from Eqs.~\ref{eq:G_3D} and \ref{eq:k_3D} as
~\cite{caupin2006cavitation,herbert2006cavitation}
\begin{align}\label{eq:p_cav_3D}
  p_\mathrm{3Dcav}^{2} = \frac{16 \pi\gamma^3}{3k_\textrm{B} T}
  \,\frac{1}{\ln(\kappa_\mathrm{3D}\,L^3\,\tau)}\,.
\end{align}
It is seen that $ p_\mathrm{3Dcav}$ depends only logarithmically and thus rather weakly
on system size $L$ and cavitation time $\tau$. Various methods for determining $\kappa_\mathrm{3D}$ have been proposed~\cite{blander1975bubble,pettersen1994experimental}.
In a previous simulation study on bulk cavitation, we determined
$\kappa_\mathrm{3D}=5 \times 10^{11}\,\mathrm{ns^{-1}nm^{-3}}$ for the SPC/E water
model~\cite{kanduc_cavitation_2020}, this value will be employed in the current study.
Using the same water model and the same methodology ensures consistency and enables a direct comparison between bulk, surface and defect cavitation rates.
It is noteworthy that CNT (as given by \autoref{eq:p_cav_3D}) for SPC/E water predicts $p_\mathrm{3Dcav}$ values from $-$80 to
$-$100 MPa for $L$ between 1~\textmu m and 1 m and $\tau=1$ s. This value is slightly higher than previous predictions, primarily
because the surface tension of SPC/E water ($\gamma_\textrm{SPC/E}\approx55$~mN/m for a
Lennard-Jones cutoff 0.9~nm~\cite{kanduc_atomistic_2017}) is lower than the experimental
value $\gamma_\textrm{exp}=72$ mN/m. Applying a correction factor of
$(\gamma_\textrm{exp}/\gamma_\textrm{SPC/E})^{3/2}\approx1.5$ on $p_\mathrm{3Dcav}$, as
suggested by \autoref{eq:p_cav_3D}, yields good agreement with previous
predictions~\cite{fisher_fracture_1948, caupin2005liquid, caupin2006cavitation,
caupin2013stability, azouzi2013coherent}. We here stick to the SCP/E water model because
it reproduces most dynamic water properties quite well.

\rev{Experimental validation of the CNT bulk cavitation prediction for water is
  challenging, as the bulk cavitation pathway is preempted by diverse surface and defect
  cavitation pathways. Nevertheless, bulk cavitation occurs in liquids with considerably
  lower surface tensions, such as hydrocarbons. With surface tensions approximately three
  times smaller than water, the predicted cavitation pressures by CNT
  (\autoref{eq:p_cav_3D}) are reduced by a factor of $3^{3/2}\approx 5$, resulting in
  cavitation pressures of $-20$ to $-30$ MPa. This range agrees well with experimentally
  measured cavitation pressures in various organic liquids~\cite{galloway1954experimental,
  apfel1977tensile, ohde1993raising, vinogradov2000boundary}, providing validation of the
CNT framework for bulk cavitation. }
%%%%%%%%%%%%%%%%%%%%%%%%%%%%%%%%%%%%%%%%%%%%%%%%%%%%%%%%%%%%%%%%%%%
\subsection{Cavitation at a smooth defect-free surface}
\begin{figure}
  \includegraphics[width=\columnwidth]{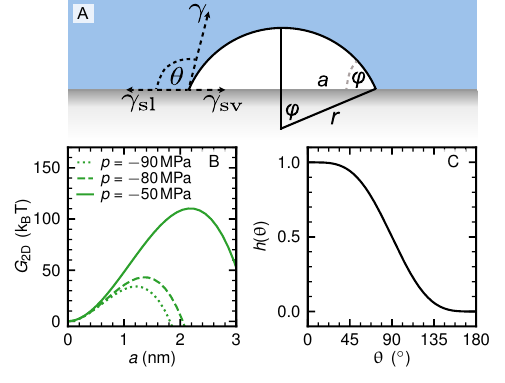}
  \caption{(A)~Schematics of bubble cavitation on a surface. Bubble contact angle
    $\varphi$ and water contact angle $\theta$ are related by $\varphi = \pi - \theta$.
    (B)~Free energy $G_\mathrm{2D}$ of a bubble on a surface with a contact angle of
    $\theta=97^\circ$ as a function of the bubble base radius $a$ from
    \autoref{eq:free_energy_2D} for various negative pressures $p$. (C)~ Geometric
  factor $h(\theta)=G_\mathrm{2D}^*/G_\mathrm{3D}^*$ in \autoref{eq:h}.}
  \label{fig:theory}
\end{figure}

To apply the CNT framework, we write the free energy of a bubble at a surface,
illustrated in \autoref{fig:theory}\,A, as
\begin{align}\label{eq:free_energy_2D}
  G_\mathrm{2D} = A_\mathrm{cap} \gamma
  + A_\mathrm{base}(\gamma_\mathrm{sv}-\gamma_\mathrm{sl})
  + pV_\mathrm{b}\,,
\end{align}
where $A_\mathrm{cap}=2\pi r^2(1 -\cos\varphi)$ is the area of the spherical bubble cap
with radius of curvature $r$, $ A_\mathrm{base}=\pi a^2$ is the base area of the bubble
with base radius $a$ and $V_b = (\pi/3)r^3(2 - 3 \cos\varphi + \cos^3\varphi)$ is the
bubble volume, $\gamma_\mathrm{sv}$ and $\gamma_\mathrm{sl}$ correspond to the
solid--vapor and solid--liquid surface tensions, respectively, and $\varphi$ is the
contact angle of the bubble. Similar to \autoref{eq:free_energy_3D}, the first two terms
account for the free energy of creating the bubble surface and the third term is the
work associated with volume expansion. We neglect the buoyancy force, as it is
irrelevant for determining the critical bubble size, which is on the nanometer scale,
and mostly important for the pinching-off of full-grown bubbles.

Using Young's equation, $\gamma_\mathrm{sv}-\gamma_\mathrm{sl} = \gamma \cos\theta$, and
the relation between the water contact angle $\theta$ and the bubble contact angle
$\varphi$, $\varphi + \theta = 180^\circ$, we show in \autoref{fig:theory}\,B the free
energy of a bubble at a hydrophobic surface characterized by a water contact angle
$\theta=97^\circ$ according to \autoref{eq:free_energy_2D} as a function of the base
radius $a$ for various pressures. The critical bubble size follows from
$\mathrm{d}G_\mathrm{2D}/\mathrm{d}a|_{a=a*} = 0$ as $a^* = -(2\gamma/p)\sin \theta$,
the corresponding radius of curvature is $r^*=-2\gamma/p$, as for homogeneous
cavitation.
%
%\begin{align}\label{eq:a_crit} a^* = -\frac{2\gamma}{p}\sin \theta \end{align}
%
The free energy barrier follows as
\begin{align}\label{eq:G_2D}
  G_\mathrm{2D}^*(\theta) = G_\mathrm{3D}^* \,h(\theta)\,,
\end{align}
which is reduced compared to the bulk result $G_\mathrm{3D}^*$ in \autoref{eq:G_3D} by
the geometric factor
\cite{debenedetti1996metastable, caupin2006cavitation, sear2007nucleation}
\begin{align}\label{eq:h}
  h(\theta)=(2-\cos \theta)
  \cos^4\left({\theta}/{2}\right)\,
\end{align}
depicted in \autoref{fig:theory}\,C. For a completely wetting surface, $\theta=0$, we
find $h(0)=1$, the bubble is detached from the surface and the bulk cavitation result is
recovered. The factor $h(\theta)$ decreases as the contact angle increases, ultimately
disappearing for $\theta=180^\circ$. The surface cavitation rate is written as
\begin{align}\label{eq:k_2D}
  k_\mathrm{2D} = A \kappa_\mathrm{2D}(\theta) \,
  e^{-\beta G^*_\mathrm{2D}(\theta)}\,
\end{align}
and is proportional to the total surface area $A$, where $\kappa_\mathrm{2D}$ is the
attempt frequency surface density, which depends on the surface type and thus on the
contact angle $\theta$.
For the total surface area of a cubic container $A=6L^2$ and neglecting edge and corner
effects, the surface cavitation pressure is given by
\begin{align}\label{eq:p_cav_2D}
  p_\mathrm{2Dcav}^2 = \frac{16 \pi\gamma^3}{3k_\textrm{B} T}
  \,\frac{h(\theta)}{\ln[\kappa_\mathrm{2D}(\theta)\, 6L^2\,\tau]}\,.
\end{align}

\begin{figure*}
  \includegraphics[width=\textwidth]{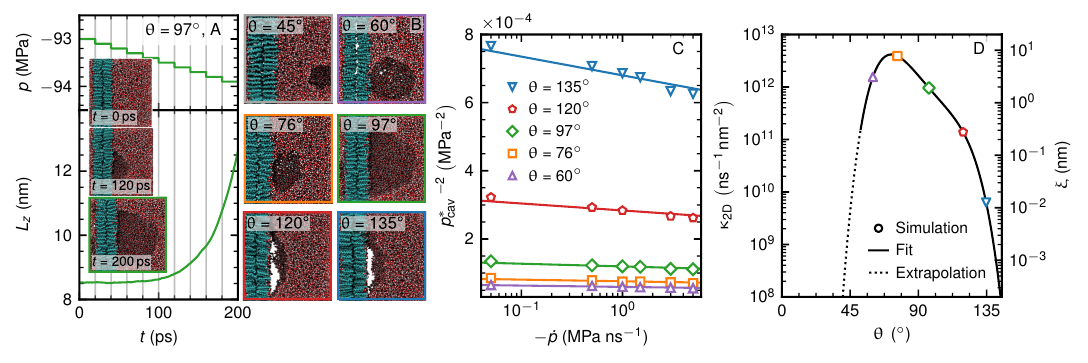}
  \caption{\label{fig:raw_results}
    (A)~Time-dependent pressure ($p$, top) and simulation box size ($L_z$, bottom)
    during a pressure ramp simulation with a pressure rate of $\dot{p}
    =-5\,\mathrm{MPa/ns}$ for a system with a surface contact angle of
    $\theta=97^\circ$. The presented trajectory corresponds to the final 200 ps prior to
    cavitation, identified as a 50\,\% increase of $L_z$ relative to its initial value
    $L_z^0=8\,\mathrm{nm}$. We find a cavitation pressure of $p_\mathrm{cav}^*
    =-93.9\,\mathrm{MPa}$. Vertical lines mark individual simulation segments at
    constant pressure. The inset shows consecutive snapshots of the cavitation event.
    (B)~Cavitation configurations with $L_z/L_z^0=3/2$ for different contact angles. We
    also present trajectory visualizations for $\theta=45^\circ$ (upper left panel)
    (multimedia available online) and $\theta=135^\circ$ (lower right panel) (multimedia
    available online). (C)~Simulation results for ${1/p_\mathrm{cav}^*}^{2}$ versus
    $\dot{p}$ for different contact angles. The lines are fits of \autoref{eq:p*cav} to
    the data, where $\kappa_\textrm{2D}$ is the only fitting parameter. (D)~Results for
    $\kappa_\textrm{2D}$ (left scale) and the ratio
    $\xi=\kappa_\mathrm{2D}/\kappa_\mathrm{3D}$ (right scale) as a function of contact
    angle $\theta$. The solid black line is a fit of Eq.~S13 in the Supplementary
    Material, section S3. \rev{For $\theta < 52^\circ$, the fit does not serve as a
  reliable extrapolation and is denoted as a broken line}.}
\end{figure*}

%We now use our previous simulation results for homogeneous
%cavitation~\cite{kanduc_cavitation_2020} and heterogeneous cavitation from this study,
%combined with CNT, to compute the cavitation pressure of a water-filled cubic container
%with size $L$.

To determine $\kappa_\textrm{2D}$ from simulations, we use a self-assembled monolayer of
alkyl molecules terminated by headgroups with adjustable electric dipole
moments~\cite{kanduc2014attraction, kanduc_atomistic_2017}. The dipole moment controls
the wetting characteristics of the surfaces. We examine six different dipole moments,
resulting in contact angles of $\theta = 45^\circ$, $60^\circ$, $76^\circ$, $97^\circ$,
$120^\circ$ and $135^\circ$, on surfaces measuring 9 nm $\times$ 10.4 nm in contact with
water. Obtaining the cavitation time from simulations at constant negative pressure is
impractical due to the excessive waiting time for cavitation to occur. Instead, we apply
negative pressures that decrease linearly over time as $p(t)=\dot p t$ with different
fixed pressure ramp rates $\dot p<0$, as used before for bulk cavitation
~\cite{kanduc_cavitation_2020} and described in the methods section.

\autoref{fig:raw_results} summarizes our simulation results. Panel A illustrates a
cavitation event on a hydrophobic surface with contact angle $\theta=97^\circ$, where a
vapor bubble forms at the surface (see inset snapshots) due to the progressively
decreasing pressure over time (upper graph). The simulation trajectory consists of a
sequence of 20-ps-long simulations (delineated by gray vertical lines), each at a
constant applied pressure that decreases in discrete steps of 0.1 MPa, which corresponds
to a mean pressure rate of $\dot{p} = -5\,\mathrm{MPa/ns}$. The formation of the bubble
is monitored by the box size normal to the surface $L_z$ (bottom graph). We define
cavitation to take place when the box size $L_z$ increases by 50\,\% compared to its
initial value $L_z^0$, which in this specific example yields a cavitation pressure
$p_\mathrm{2Dcav}^*=-93.9\,\mathrm{MPa}$. It is important to note that the cavitation
pressure in the pressure-ramp ensemble $p_\mathrm{2Dcav}^*$ is different from the
cavitation pressure $p_{\mathrm{2Dcav}}$ in the constant-pressure ensemble given in
\autoref{eq:p_cav_2D}, which normal pressure will be determined later.

Representative snapshots of cavitation for $L_z/L_z^0=3/2$ at surfaces with different
contact angles are presented in \autoref{fig:raw_results}\,B. We also provide trajectory
visualizations for $\theta=45^\circ$ (upper left panel) (multimedia available online)
and $\theta=135^\circ$ (lower right panel) (multimedia available online). \autoref{fig:raw_results}\,C shows $1/{p^*_{\rm cav}}^2$
versus the negative pressure-ramp rate $-\dot{p}$. Each data point represents an average
over 10 independent simulations.
The solid lines are fits of the kinetic theory (\autoref{eq:p*cav} and using $G^*=G^*_\mathrm{2D}$), assuming pure surface cavitation which yield the value of $\kappa_\textrm{2D}(\theta)$ for each $\theta$.
For the lowest contact angle of 45$^\circ$, bubble cavitation occurs in the bulk. Therefore,
this surface is not included in the analysis in \autoref{fig:raw_results}\,C. The
resulting values for $\kappa_\textrm{2D}$ (symbols) are shown as a function of the
surface contact angle $\theta$ in \autoref{fig:raw_results}D. We also denote the
characteristic length $\xi=\kappa_\textrm{2D}/\kappa_\textrm{3D}$ by the scale on the
right, which represents the water slab height for which the surface cavitation attempt
frequency $\kappa_\textrm{2D}$ equals the bulk attempt frequency $\kappa_\textrm{3D}$.
It can be seen that $\xi$ exhibits a maximum around $\theta \approx 75^\circ$ and goes
dramatically down with increasing contact angle from approximately $10$ to $10^{-2}$ nm.
This suggests that a hydrophilic surface produces more cavitation nuclei in a given time
than a hydrophobic surface, which is interesting since it contrasts with the pronounced
density fluctuations at hydrophobic surfaces \cite{AmishPatel} (alternatively, it could
reflect inaccuracies of the bulk and surface-barrier-free energy expressions).
Clearly, this variation of $\kappa_\textrm{2D}$ is subdominant compared to the variation
of the exponential barrier energy in \autoref{eq:k_2D}, but nevertheless has to be
included for a quantitative comparison of surface and bulk cavitation kinetics. For our
subsequent analyses, we fit our simulation data for the logarithm of
$\kappa_\mathrm{2D}$ in \autoref{fig:raw_results}D to a \rev{polynomial} shown as a
solid line, as detailed in the Supplementary Material, section S3. \rev{While polynomial
  fits are suitable for interpolation, they are less reliable for extrapolation. We
  therefore trust the polynomial fit only down to a constant angle extrapolation intervall
  $60^\circ - \theta^* = 8^\circ$ which is half the angle distance between the two lower
  data points $76^\circ - 60^\circ = 16^\circ$. Therefore the fit for $\theta < \theta^* =
52^\circ$ is shown as a broken line.}

\begin{figure}
  \includegraphics[width=\columnwidth]{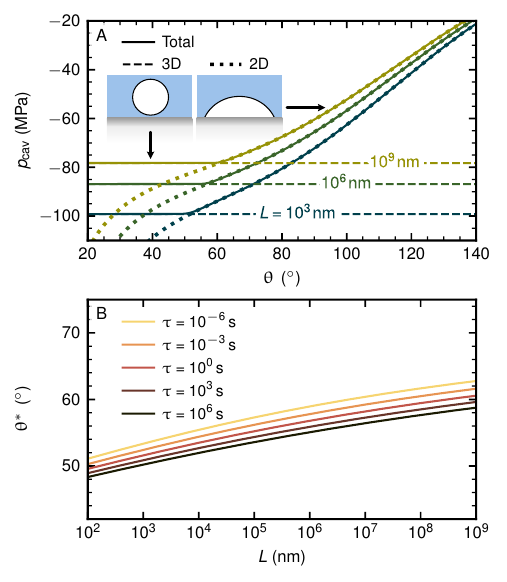}
  \caption{\label{fig:cav_pressure_theta}
    (A)~Cavitation pressure $p_\mathrm{cav}$ of a water-filled cubic container as a
    function of the contact angle $\theta$ of the defect-free inner walls, for a fixed
    waiting time of $\tau=1\,\mathrm{s}$, computed by numerically inverting
    \autoref{eq:k_tot} for $N_\textrm{def}=0$ (solid lines). Different colors represent
    different cube sizes $L$. Dotted lines show the 2D cavitation pressure according to
    \autoref{eq:p_cav_2D}, while dashed horizontal lines denote the 3D cavitation pressure based on
    \autoref{eq:p_cav_3D}. (B)~ Crossover contact angle $\theta^*$, where
    $k_\mathrm{2D}=k_\mathrm{3D}$, as a function of container size $L$ for different
  waiting times $\tau$, according to \autoref{eq:L_theta_star}.}
\end{figure}

%%%%%%%%%%%%%%%%%%%%%%%%%%%%%%%%%%%%%%%%%%%%%%%%%%%%%%%%%%%%%%%%%%%
\subsection{Kinetic competition between cavitation in bulk and at a defect-free surface}

In the absence of defects, we set $N_\mathrm{def}=0$ in \autoref{eq:k_tot}. The
cavitation pressure $p_\mathrm{cav}$ cannot be derived analytically, therefore, we solve
\autoref{eq:k_tot} numerically and plot $p_\mathrm{cav}$ in
\autoref{fig:cav_pressure_theta}\,A (solid lines) against the surface contact angle
$\theta$ for different cube sizes $L$ and a fixed waiting time of
$\tau=k_\mathrm{tot}^{-1}=1\,\mathrm{s}$. For hydrophilic surfaces with small $\theta$,
$p_\mathrm{cav}$ is dominated by bulk cavitation (thus, $k_\textrm{tot}\approx
k_\textrm{3D}$) and becomes independent of $\theta$, as described by
\autoref{eq:p_cav_3D} and represented by the horizontal dashed lines. For larger contact
angles $\theta$, the cavitation pressure is well described by the surface prediction
(\autoref{eq:p_cav_2D}), shown as dotted lines, which indicates that cavitation for large
$\theta$ predominantly occurs at surfaces.

It is seen that cavitation shifts from the bulk (3D) to the surface (2D) pathway rather
abruptly at a distinct crossover contact angle $\theta^*$, which marks the onset
of surface cavitation. Thus, $\theta^*$ is identified as the intersection of the bulk
prediction (dashed lines) and the surface prediction (dotted lines) in
\autoref{fig:cav_pressure_theta}\,A. In Supplementary Material, section S4, we present
results for a range of different waiting times $\tau$ from $10^{-3}$ to $10^3$~s,
showing the same sharp transition between the surface-dominated and the bulk-dominated
regimes. By equating \autoref{eq:p_cav_3D} and \autoref{eq:p_cav_2D}, we obtain an exact
expression for the critical system size $L^*$ at which the transition takes place
\begin{align}\label{eq:L_theta_star}
  L^* = \left [
    6 \tau \kappa_\mathrm{2D}(\theta)
  (\kappa_\mathrm{3D}\tau )^{-h(\theta)}\right ] ^{1/[3h(\theta)-2]}\,,
\end{align}
whose inverse (i.e., $\theta^*$ as a function of $L$) is presented in
\autoref{fig:cav_pressure_theta}\,B for various cavitation times $\tau$. It transpires
that $\theta^*$ varies within a narrow window between $50^\circ$ and $60^\circ$ for
order-of-magnitude variations of system size $L$ and cavitation time $\tau$.

This weak dependence of $\theta^*$ on $L$ and $\tau$ can be understood from the
analytical inverse of \autoref{eq:L_theta_star} valid for small $\theta$,
\begin{equation}
  \theta^* \approx \left\{ \frac{16 \ln[L/(6 \xi)]}
  {3 \ln (\kappa_\mathrm{3D} L^3 \tau)} \right\}^{1/4}\,,
\end{equation}
see Supplementary Material, section S5 for the derivation. The critical contact angle
depends as a fourth root on the logarithm of $L$ and $\tau$, which is an
extremely weakly varying function.

We conclude that surface cavitation occurs if the contact angle of the confining
surfaces exceeds the threshold value $\theta^*$, which is largely independent of system
size and waiting time. Although surfaces with contact angles below the threshold (i.e.,
for $\theta<\theta^*$) do have a lower cavitation free energy barrier compared to bulk
cavitation, as predicted by \autoref{eq:G_2D}, they exhibit a smaller cavitation rate
due to the different attempt frequencies at a surface and in bulk, as quantified by the
length scale $\xi$. This renders surface cavitation on hydrophilic surfaces less likely
than bulk cavitation.

\subsection{Cavitation at surface defects}
\begin{figure*}
  \includegraphics[width=\textwidth]{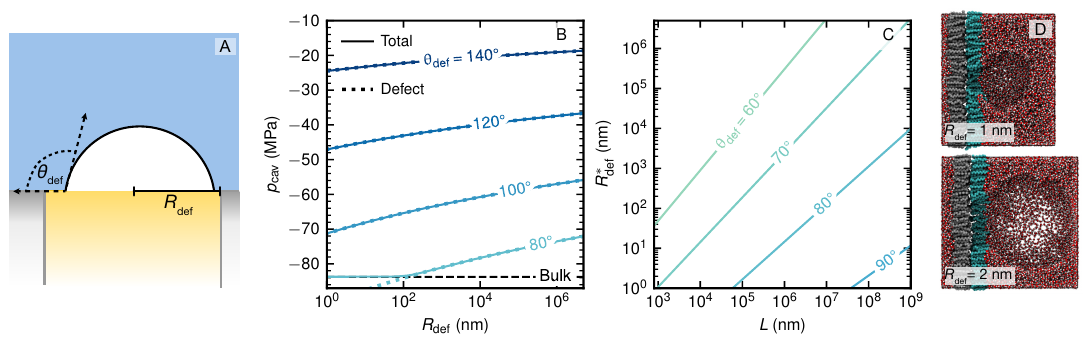}
  \caption{\label{fig:cav_pressure_def}
    (A)~Illustration of a cavitation bubble on a circular surface defect with radius
    $R_\textrm{def}$ and contact angle $\theta_\textrm{def}$ on a flat surface of
    contact angle $\theta$. (B)~Cavitation pressure of a water-filled container with a
    single surface defect $N_\textrm{def}=1$ as a function of $R_\textrm{def}$ for
    different $\theta_\textrm{def}$. Solid lines are obtained by numerically solving
    \autoref{eq:k_tot} for a waiting time of $\tau=1\,\mathrm{s}$ and a container size
    of $L=1\,\mathrm{cm}$ (volume of $1\,\mathrm{mL}$). Dotted lines show the limiting
    pressure for a large defect, given by~\autoref{eq:p_cav_2D_defect}, whereas the horizontal
    dashed line shows the bulk cavitation pressure \autoref{eq:p_cav_3D}. (C)~Crossover
    defect radius $R_\mathrm{def}^*$ as a function of container size $L$ for a fixed
    waiting time of $\tau=1\,\mathrm{s}$ from \autoref{eq:R_def}. (D)~Simulation
    snapshots for box expansions of $L_z/L_z^0=3/2$ after cavitation in systems with a
    surface contact angle of $97^\circ$ and cylindrical hydrophobic pits, from
    pressure-ramp simulations at a rate of $\dot{p} = -5\,\mathrm{MPa/ns}$. The pits
    have radii of $R_\textrm{def} \approx 1$ nm (upper panel) (multimedia available
    online) and $2~\mathrm{nm}$ (lower panel) (multimedia available online).}
\end{figure*}

Real-world surfaces are rich in surface defects such as vacancies, disordered regions,
variations in material composition and adsorbed organic or inorganic
species~\cite{siretanu2016atomic}. We model defects as circular patches with radius
$R_\mathrm{def}$ and contact angle $\theta_\mathrm{def}$, as illustrated in
\autoref{fig:cav_pressure_def}\,A. We assume the defects to be larger than the critical
bubble base radius $a^*$ and much smaller than the system size $L$, $a^*\ll
R_\mathrm{def}\ll L$.
With these assumptions, the single defect cavitation rate in \autoref{eq:k_tot} is given
in analogy to \autoref{eq:k_2D} as
\begin{align}\label{eq:k_def}
  k_\mathrm{def} = \kappa_\mathrm{2D}(\theta_\mathrm{def}) A_\mathrm{def}\,
  e^{-\beta G^*_\mathrm{2D}(\theta_\mathrm{def})}\, ,
\end{align}
with the defect surface area given by $A_\mathrm{def}=\pi R_\mathrm{def}^2$ and the
attempt frequency density $\kappa_\mathrm{2D}$ and the free energy barrier
$G^*_\mathrm{2D}$ determined by the defect contact angle $\theta_\mathrm{def}$.
The defect cavitation pressure follows from
$\tau^{-1}= N_\mathrm{def} k_\mathrm{def}$ by using
\autoref{eq:k_def} and
\autoref{eq:G_2D}
as
\begin{align}\label{eq:p_cav_2D_defect}
  p_\textrm{defcav}^2 = \frac{16\pi \gamma^3}{3k_\textrm{B} T}\,
  \frac{h (\theta_\mathrm{def}) }
  {\ln[ \kappa_\mathrm{2D}(\theta_\mathrm{def})\, \pi R_\mathrm{def}^2\,\tau N_\mathrm{def} ]}\,.
\end{align}
The total cavitation rate in \autoref{eq:k_tot} describes the kinetic competition
between cavitation in bulk, at surfaces and at defects. The cavitation pressure
resulting from this competition is determined by solving \autoref{eq:k_tot} in
conjunction with Eqs. \ref{eq:k_3D}, \ref{eq:k_2D}, \ref{eq:k_def} for a given set of
parameters $L$, $\tau$, $\theta$, $N_\mathrm{def}$, $\theta_\mathrm{def}$ and $R_\mathrm{def}$.

For the following discussion, we assume that the surface is sufficiently hydrophilic so
that $k_\mathrm{2D}\ll k_\mathrm{3D}$ and cavitation at the defect-free surface parts
can be neglected. The solid lines in \autoref{fig:cav_pressure_def}B show the cavitation
pressure in a 1~mL cube of water ($L=1$~cm) in the presence of a single surface defect,
$N_\mathrm{def}=1$, as a function of the defect radius $R_\mathrm{def}$ for different
defect contact angles $\theta_\textrm{def}$. For a given $\theta_\textrm{def}$, a
well-defined crossover radius $R^*_\mathrm{def}$ emerges at which the system
transitions from bulk cavitation for small $R_\mathrm{def}$, where the bulk prediction
\autoref{eq:p_cav_3D} (dashed line) is valid, to defect cavitation for large
$R_\mathrm{def}$, where the prediction in \autoref{eq:p_cav_2D_defect} (dotted lines) is
valid.
The crossover radius $R^*_\mathrm{def}$ grows as $\theta_\textrm{def}$ approaches
$\theta^*$ from above. Conversely, for increasing $\theta_\textrm{def}$ the crossover
radius $R^*_\mathrm{def}$ approaches sub-nanometer values (where our continuum model
will eventually break down). The results in \autoref{fig:cav_pressure_def}B show that
cavitation is dominated by defects with a radius larger than a critical value
$R^*_\mathrm{def}$ and occurs at a rate $N_\textrm{def} k_\textrm{def}$; even a single
hydrophobic defect can dominate the cavitation kinetics in macroscopic systems.

The crossover radius $R^*_\mathrm{def}$ between the cavitation pressures for the bulk
and defect pathways follows from equating Eqs. \ref{eq:p_cav_3D} and
\ref{eq:p_cav_2D_defect} as
\begin{align}\label{eq:R_def}
  R_\mathrm{def}^{*2} = \frac{
  \left(\kappa_\mathrm{3D} L^3 \tau \right)^{h(\theta_\mathrm{def})}}{
  \pi \kappa_\mathrm{2D}(\theta_\mathrm{def})\tau N_\mathrm{def} }\,.
\end{align}
In \autoref{fig:cav_pressure_def}\,C we show $R_\mathrm{def}^*$ as a function of system
size $L$ for a waiting time of $\tau=1\,\mathrm{s}$ and a single defect ($N_\mathrm{def}=1$)
for four different defect contact angles. As the system size increases, a larger defect
is required to induce defect cavitation. Conversely, for larger defect contact angle
$\theta_\textrm{def}$, the crossover radius $R_\textrm{def}^*$ decreases and reaches the
nanometer scale even for macroscopically large systems. For a surface with multiple defects,
$N_\mathrm{def}>1$, the required defect radius $R_\mathrm{def}^*$ is expected to decrease proportionally to $\sim N_\mathrm{def}^{-1/2}$, as predicted by \autoref{eq:R_def}.

Finally, we check our theoretical predictions by explicit simulations of a system
featuring one defect. We choose a mildly hydrophobic surface with a contact angle of
$\theta=97^\circ$ on which we create one circular defect by removing surface molecules
within a radius of $R_\mathrm{def}$ on one monolayer, thereby forming a hydrophobic pit
with a depth of 1 nm. Water does not enter the pit because of its hydrophobic interior,
creating a vapor bubble, corresponding to a defect contact angle of
$\theta_\textrm{def}\approx180^\circ$. To avoid cavitation at the opposite monolayer, we
make it hydrophilic with a contact angle of $42^\circ$. Using the pressure ramp
protocol, we consistently observe bubble formation at the hydrophobic pit, as shown in
the snapshots in \autoref{fig:cav_pressure_def}\,D, despite the rest of the surface
being mildly hydrophobic. Trajectory visualizations for $R_\mathrm{def}=1$~nm (upper panel) (multimedia
available online) and $R_\mathrm{def}=2$~nm (lower panel) (multimedia available online) defects are
provided as well. Using a pressure rate of $\dot{p} = -5\,\mathrm{MPa/ns}$, a
defect-free surface induces surface cavitation at $p^*_\mathrm{cav}=-95 \pm 1$ MPa, as
seen in \autoref{fig:raw_results}\,C. In contrast, the systems with surface defects of 1
and 2 nm radii cavitate already at $p_\mathrm{cav}^*=-66$ and $-45 \pm 2 $ MPa,
respectively. These results qualitatively agree with our predictions in
\autoref{fig:cav_pressure_def}\,B that even nanometer-sized surface defects can act as
cavitation nuclei. \rev{The influence of defects on cavitation is studied in more detail
in our follow-up study \cite{Sako2024pit}.}

%% file: main/sections/conclusion.tex
\section{Conclusion}
By combining MD simulations with the CNT framework, we explore the kinetic competition
\rev{among the three fundamental pathways of cavitation in water under negative pressure:} bulk
cavitation, surface cavitation and defect-induced cavitation. For defect-free surfaces,
the prevalent cavitation pathway shifts abruptly from bulk to surfaces when
the surface contact angle increases above a threshold of $\theta^* \approx
50^\circ$ to $60^\circ$, depending slightly on system size and waiting time. The typical
cavitation pressure decreases in magnitude from about $p_\mathrm{cav} \approx -100$ MPa
for bulk to about $p_\mathrm{cav} \approx -30$ MPa for hydrophobic surface cavitation.
Even the presence of a single nanoscopic hydrophobic surface defect dominates the
cavitation kinetics and substantially raises the cavitation pressure, depending
on the defect contact angle and defect size.
Since even with the most advanced fabrication techniques surface defects are
unavoidable and experimental cavitation pressures for typical surface
materials are far above the predicted value for bulk cavitation and low cavitation
pressures around $p_\mathrm{cav} \approx-100$ MPa are only achieved in micro-sized
defect-free quartz cavities~\cite{zheng_liquids_1991, alvarenga1993elastic,
azouzi_coherent_2013}.

%We have considered contamination-free water, i
\rev{While hydrophobic aggregates reduce the water stability against
cavitation, individual dispersed molecules generally have a minimal
effect~\cite{zhou2019molecular, kanduc_cavitation_2020}. Viscosifying agents, such as
polymers, slow down cavitation kinetics, by
influencing the prefactor $\kappa$ in Eqs.~\eqref{eq:k_3D} and \eqref{eq:k_2D} without
altering the exponential term.} However, surfaces with contact angles exceeding the
``Berg limit'' of approximately 65°---coincidentally near the crossover contact
angle---are prone to adhesion of amphiphilic molecules~\cite{vogler1998structure,
rosenhahn2010role, sako2023conditions, kanduc2024understanding}. This adhesion can
render such surfaces more hydrophilic and coat surface defects~\cite{sako2023conditions},
thereby suppressing surface cavitation. \rev{This mechanism might explain the prevention
of embolisms through lipid adsorption in tree sap~\cite{schenk2015nanobubbles} or the
increased cavitation stability observed when amphiphilic polymers are added to
water~\cite{brujan_cavitation_2008, gruzdkov_cavitation_2008}. Exploring the effect of molecular
adsorption on defects and cavitation is an intriguing
question, which we aim to pursue in upcoming studies.}

In this study, we focused on pressure-induced cavitation in water at constant
temperature. However, the methods developed here can be extended to other liquids and
applied to cavitation in superheated liquids or nucleation in supersaturated solutions.
In such systems, surface and defect nucleation are expected to play similarly critical
roles. 

%% file: main/sections/supplementary.tex
\section{Supplementary Material}
The supplementary material contains additional information on fitting the kinetic
prefactor, the cavitation pressure for various waiting times $\boldsymbol{\tau}$, the
relation between surface polarity $\alpha$ and contact angle of the SAM surface.

%% file: main/sections/acknowledgments.tex
\begin{acknowledgments}
    M.K.\ acknowledges financial support from the Slovenian Research and Innovation
    Agency ARIS (contracts P1-0055 and J1-4382).
\end{acknowledgments}

%% file: main/sections/data_availability.tex
\section*{Data Availability Statement}
All simulation input files as well as Python functions to fit $\kappa_\mathrm{2D}$ and
perform further analysis from the simulations are available on
\href{https://zenodo.org/doi/10.5281/zenodo.14236547}{Zenodo}.

%% file: si/sections/polarity_angle.tex
\section{Relation between surface polarity $\boldsymbol{\alpha}$ and contact angle
$\boldsymbol{\theta}$}
\label{sec:polarity_tension}

\begin{SCfigure}
  \centering
  \includegraphics{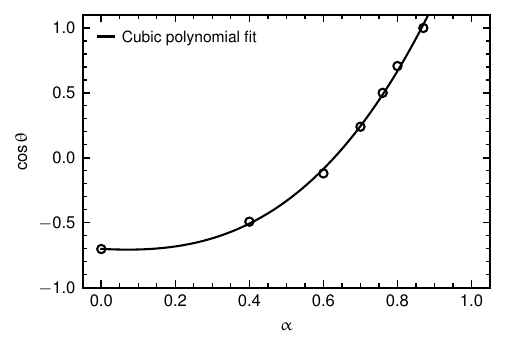}
  \caption{\label{fig:contact_angle_fit}Cosine of the contact angle $\theta$ as a
    function of the SAM's surface polarity $\alpha$. Data is taken from
    Ref.~\citenum{kanduc_atomistic_2017}. The solid line is the fit of
  \autoref{eq:theta_to_alpha} to the data points.\vspace{13em}}
\end{SCfigure}

The used model surface, resembling SAMs with tunable partial charges in the hydroxyl
groups, has been thoroughly studied elsewhere~\cite{kanduc_atomistic_2017}.
\autoref{fig:contact_angle_fit} shows the cosine of the contact angle $\cos\theta$ as a function of the
scaling factor $\alpha$ of the hydroxyl group's partial charges. For $\alpha=0$ (i.e.,
the partial charges of the hydroxyl groups are 0), the surface is nonpolar, resulting in
a contact angle of 135°. For $\alpha>0.86$, the surface exhibit complete wetting, with
$\theta=0$. We fit the simulation data to
\begin{align}\label{eq:theta_to_alpha}
  \cos\theta = c_0 + c_1 \alpha + c_2 \alpha^2 + c_3 \alpha^3\,,
\end{align}
yielding the coefficients: $c_0 = -0.70$, $c_1 = -0.09$, $c_2 = 0.65$, and $c_3 = 1.99$.
The fit is shown as a solid line in \autoref{fig:contact_angle_fit}. This fit was used
to determine the partial charges corresponding to a contact angle of $\theta=60^\circ$,
which had not been previously determined.

%% file: si/sections/fitting_prefactor.tex
\section{Fitting the kinetic prefactor}
\label{sec:fitting}
Using classical nucleation theory, one can derive the free energy barriers as given by
Eqs.~3 and~7. However, the kinetic prefactor is governed by hydrodynamic and dissipation
effects, making it difficult to assess analytically. Atomistic simulations provide a way
to extract this factor. A straightforward approach might be to simulate a system at a
constant negative pressure and track bubble formation. Unfortunately, this approach
makes it challenging to observe cavity formation within practical simulation times,
particularly without prior knowledge on the cavitation pressure. Therefore, we apply a
time-dependent pressure ramp in the form
\begin{align}
  p(t) = \dot{p}\,t\,,
\end{align}
to sample these rare events by solving time-dependent rate equations.
Under this time-dependent pressure, the cavitation rate also becomes time-dependent and
can be, in the Kramers framework, approximately expressed as
\begin{align}
  k(t) = k_0 e^{-\beta G^*(t)}\,.
\end{align}
Note that for clarity we leave out the subscripts 3D or 2D since the equations are valid
for both. The probability $f(t)$ that the system has not yet crossed the barrier obeys a
first-order rate equation
\begin{align}
  \dot f(t) = -k(t)\,f(t)\,,
\end{align}
with the formal solution
\begin{align}
  f(t) = e^{-k_0 I(t)}\,,
\end{align}
where $I(t)$ is given by
\begin{align}\label{eq:kernel}
  I(t) = \int_0^t e^{-\beta G^*(t')} \mathrm{d}t'\,.
\end{align}
In our pressure-ramp simulations, the free-energy barrier (given by Eqs.~3 or 7)
decreases inversely with the square of time as
\begin{align}
  \label{eq:Gt}
  \beta \Delta G^*(t) = \left(\frac{\zeta}{t}\right)^2\,,
\end{align}
with the time constants of the barrier $\zeta$ for homogeneous (3D) and heterogeneous
(2D) cavitation given by
\begin{align}
  \left(\zeta_\mathrm{3D}\right)^2 &= \frac{16 \pi }{3}\frac{\gamma^3}{\dot{p}^2}\,,\\
  \zeta_\mathrm{2D} &= \zeta_\mathrm{3D} h(\theta) \,.
\end{align}
With the time-dependent free energy barrier given by Eq.~\ref{eq:Gt}, \ref{eq:kernel}
has the closed-form solution
\begin{align}
  I(t) = t \exp\left[-\left(\frac{\zeta}{t}\right)^2\right]-\sqrt{\pi}\,\zeta\,
  \mathrm{erfc} \left(\frac{\zeta}{t}\right)\,,
\end{align}
where $\mathrm{erfc}(x)$ is the complementary error function. The mean cavitation time
in the pressure-ramp protocol follows as
\begin{align}
  \tau = -\int_0^\infty t  \dot f(t) \mathrm{d}t = \int_0^\infty f(t) \mathrm{d}t\,.
\end{align}
Defining the dynamic cavitation pressure in the pressure-ramp protocol as
$p_\mathrm{cav}^* = \dot{p} \,\tau$, one finds
\begin{align}\label{eq:p*cav}
  p_\mathrm{cav}^* = \dot{p} \int_0^\infty e^{-k_0 I(t)} \mathrm{d}t\,.
\end{align}
During a simulation, we apply a negative pressure rate $\dot{p} < 0$ (i.e., the pressure
is decreasing) and identify $p^*_\mathrm{cav}$, the pressure at which cavitation occurs.
Simulating several rates allows us to fit \autoref{eq:p*cav}, where the kinetic
prefactor $k_0$ is the only fitting parameter.

%% file: si/sections/kappa_0_2D_exponent.tex
\section{Fitting $\boldsymbol{\kappa_\mathrm{2D}(\theta)}$}
\label{sec:kappa_0_2D_fit}

\begin{figure}[h]
  \includegraphics{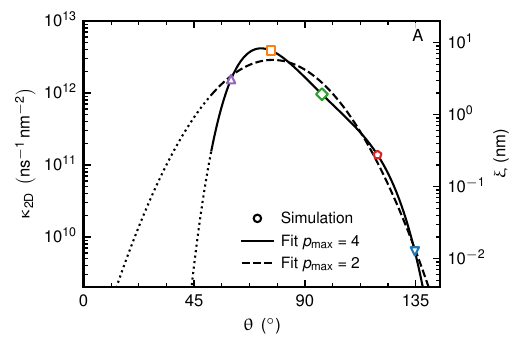}
  \includegraphics{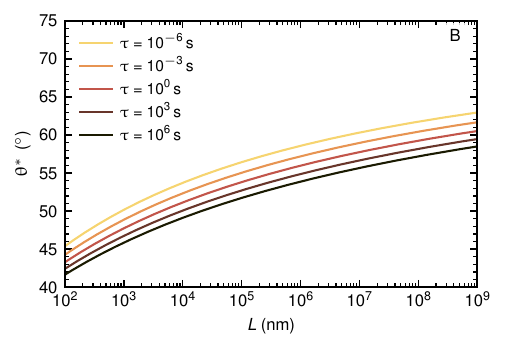}
  \caption{\label{fig:kappa_0_2D_fit}
    (A)~Fit of \autoref{eq:kappa_2D_theta_4} (solid line, as in Fig.~3D in the main
    text) and \autoref{eq:kappa_2D_theta_2} (dashed line) to the $\kappa_\mathrm{2D}$
    data points from simulations. \rev{As in the main text the fits for $\theta <
    52^\circ$ are shown as dotted lines.} (B) Crossover contact angle $\theta^*$ as a
    function of container size $L$ for different waiting times $\tau$. Similar to
    Fig.~4B, but here based on the fit of \autoref{eq:kappa_2D_theta_2} in Eq.~11.}
\end{figure}

To fit the data for $\kappa_\mathrm{2D}(\theta)$ in Fig.~3D in the main text, we use the
following fitting function,
\begin{align}\label{eq:kappa_2D_theta_4}
  \kappa_\mathrm{2D}(\theta) =
  \kappa_\mathrm{2D, 0} \, \mathrm{exp}
  \left[
    \sum_{p=1}^4 (\theta/\theta_p) ^ p
  \right] \,.
\end{align}
The fit yields $\kappa_\mathrm{2D, 0}=1.93\times 10^{-35}\,\mathrm{ns^{-1} nm^{-2}}$,
$\theta_1=0.21^\circ$, $\theta_2=3.70^\circ$, $\theta_3=12.61^\circ$, and
$\theta_4=29.77^\circ$. The resulting fit is shown as a solid line in Figs.~2D and
\ref{fig:kappa_0_2D_fit}A and was used in the subsequent analysis in the main text.

We also examine the implications of using a fitting function with a quadratic polynomial
in the exponent for the subsequent analysis, given by
\begin{align}\label{eq:kappa_2D_theta_2}
  \kappa_\mathrm{2D}(\theta) =
  \kappa_\mathrm{2D, 0} \, \mathrm{exp}
  \left[
    \sum_{p=1}^2 (\theta/\theta_p) ^ p
  \right] \,.
\end{align}
This fit, plotted as a dashed black line in \autoref{fig:kappa_0_2D_fit}A does not match
the data as well as the fourth-order polynomial fit. More critically, when we compute
the resulting crossover contact angle $\theta^*$ (using Eq.~11), shown in
\autoref{fig:kappa_0_2D_fit}B, this quadratic fit predicts $\theta^*$ below $45^\circ$
for small box sizes $L$. This result is in contradiction with the simulation outcome for
a contact angle of $45^\circ$, where cavitation in the bulk water phase occurs.
Therefore, a fit limited to second order does not adequately extrapolate to small
contact angles.

%% file: si/sections/cavitation_pressure_tau.tex
\section{Cavitation pressure for various waiting times $\boldsymbol{\tau}$}
\label{sec:cav_pressure_tau}

\begin{figure}[h!]
  \includegraphics{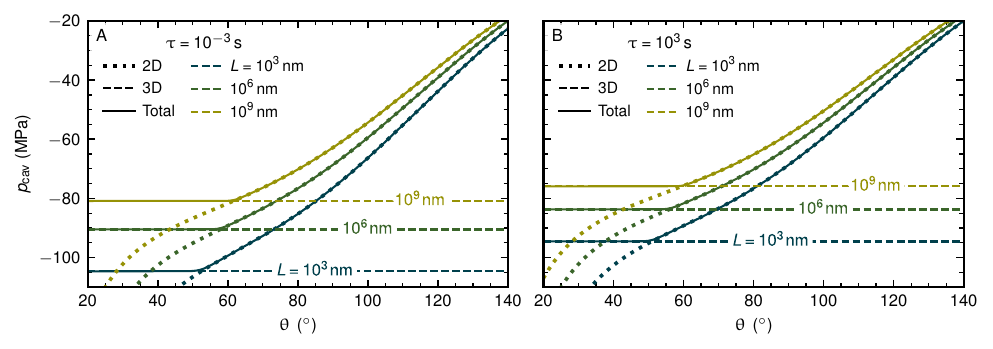}
  \caption{\label{fig:cav_pressure_theta}
    Cavitation pressure $p_\mathrm{cav}$ as a function of contact angle $\theta$,
    similar to Fig.~4A in the main text, here shown for waiting times of (A)
    $\tau=1\,\mathrm{ms}$ and (B) $\tau=1000\,\mathrm{s}\approx 17\,\mathrm{min}$.}
\end{figure}

In \autoref{fig:cav_pressure_theta}, we present the cavitation pressure as a function of
contact angle, similar to Fig.~4A in the main text, but for considerably shorter
(\autoref{fig:cav_pressure_theta}\,A) and longer (\autoref{fig:cav_pressure_theta}\,B)
waiting times. We find that increasing $\tau$ leads to a less negative cavitation
pressure $p_\mathrm{cav}$ and a lower crossover contact angle $\theta^*$ at which the
cavitation transits from bulk to surface. The latter trend is also
quantified in Fig.~4B.

%% file: si/sections/theta_star.tex
\section{Analytic estimate for the crossover contact angle $\boldsymbol{\theta^*}$}
\label{sec:theta_star}

We derive an approximate expression for the crossover contact angle $\theta^*$, which
follows from Eq.~11 in the main text. Taking the logarithm of both sides of Eq.~11 gives
\begin{equation}
  \ln L=\frac{1}{3h(\theta^*)-2}\left[\ln(6\kappa_\textrm{2D}\tau)-
    h(\theta^*)\,\ln(\kappa_\textrm{3D}\tau)
  \right]
\end{equation}
from which $h(\theta^*)$ follows as
\begin{equation}
  h(\theta^*)=\frac{\ln(6\kappa_\textrm{2D}L^2\tau)}{\ln(\kappa_\textrm{3D}L^3\tau)}\,.
  \label{eq:h_ratio}
\end{equation}
By expressing $\kappa_\textrm{2D}$ in terms of
$\xi=\kappa_\textrm{2D}/\kappa_\textrm{3D}$, we obtain
\begin{equation}
  \label{eq:h_ratio2}
  h(\theta^*)=1+\frac{\ln(6\xi/L)}{\ln(\kappa_\textrm{3D}L^3\tau)}\,.
\end{equation}
%
%The logarithm in the denominator can be linked to Eq.~3 in the main text, which
%describes the free energy barrier that is surmounted within the time $\tau$, expressed
%as \begin{equation} \beta     W_{kT}^\textrm{max}=\ln(\kappa_\textrm{3D}L^3\tau)\,.
%\end{equation} Here, $ W_{kT}^\textrm{max}$ can be interpreted as the maximal
%``effective'' thermal fluctuation expected in the system. In our two-state model, this
%value involves kinetic contributions related to the diffusion process for bubble
%formation, as described in the Kramers formalism~\cite{hanggi1990reaction}. To estimate
%the scale of these  fluctuations, consider two limiting examples: In a small,
%mesoscopic volume of 1~\textmu m$^3$ of water observed over 1~\textmu s, the largest
%expected effective thermal fluctuation is  around $55\,k_\textrm{B}T$, based on our
%estimate for $\kappa_\textrm{3D}$. In contrast, in a large, macroscopic volume of 1
%liter observed over 1 second, the largest fluctuation is likely around
%100~$k_\textrm{B}T$. Thus, in most practical scenarios, the largest fluctuation is
%expected to fall between $55$ and 100~$k_\textrm{B}T$. This relatively narrow energy
%scale provides insight into the barrier size that mesoscopic and macroscopic systems
%can realistically overcome.

Next, we perform a Taylor expansion of the geometric function $h(\theta^*)$, given by
Eq.~8, for small angles, which approximates as $h(\theta)\simeq
1-\tfrac{3}{16}\theta^4$. Substituting this into Eq.~\ref{eq:h_ratio2} leads to the
final expression
\begin{equation}
  {\theta^*}\approx\left\{\frac{16\, \ln[(L/(6\xi)]}{3\,\ln(\kappa_\textrm{3D}L^3\tau)}\right\}^{1/4}\,,
  \end{equation}
  which is Eq.~12 in the main text.